# Compensation of Reactive Power in Grid-Connected Solar PV Array System Using STATCOM and Fixed Capacitor Bank

CH Venkata Ramesh[1], A Manjunatha[2]

[1] Assistant Professor, Department of E&EE, NMIT, Bangalore.
[2]Professor, Department of E&EE, SKIT, Bangalore and affiliated to Visvesvaraya Technological University, Belagavi, Karnataka, India

[1]cvram256@gmail.com, [2] manjuprinci@gmail.com

**Abstract** - *In this article, we propose reactive compensation for the PV integrated grid system using a STATCOM and a fixed capacitor bank. This paper presents a design calculation for a PV integrated grid system with a fixed capacitor and STATCOM. The proposed system is simulated and tested using the MATLAB Simulink software package. The suggested system has been evaluated under a variety of operating circumstances, including changing solar PV array irradiance and changing reactive load power. Detailed simulation and comparisons between the fixed capacitor and STATCOM represented.*

## I. INTRODUCTION

Because of the world's growing infrastructure and population, electric power and energy consumption is increasing every day. These energy demands and power demands can be met by constructing new conventional power plants such as thermal and hydropower stations. There is a problem in the conventional power plant, such as it produces greenhouse gas, occupies a large area for construction, and needs a continuous supply of raw materials to produce electrical energy. A non-conventional power plant or renewable power plant is developed to overcome the above-stated problems [1-2]. The solar PV system is attracting the current electrical energy generation sector due to ease of installation, ease in controlling of the plant; the Manufacturing price of the PV panel is very low when compared to another renewable energy plant [3-4].

A combination of serial and parallel is commonly used in PV arrays to create photovoltaic PV array plants. By connecting the PV panels in series, the PV plant voltage will be raised, and the current rating of the solar PV power plant may be enhanced by connecting the PV panels in parallel. After making series and parallel connections, it needs to be connected to the load or connected to the grid. Generally, grid integration of the solar PV array can be done in two ways one is single stage connection, and another one is double stage connection. In a single-stage connection, the grid is directly integrated into the PV array via a grid-tied inverter with some grid control logic, and this method of connection is only useful when the terminal voltage of the PV array is within the specified level. If the

PV array's terminal voltage is low, a double-stage connection is recommended. In double stage connection, PV array voltage is step up to required standard level using direct current setup converter, and then direct current step-up converter output is integrated to the grid via a grid-tied inverter to share the power. When solar PV array voltage is very high (high than standard value), double stage connection system should use step down direct current converter between grid-tied inverter and solar PV array [5-7].

The most important thing in the solar PV integrated grid system is reactive power compensation. The real power only is supplied to the load by using a PV array system, and also excess real power is shared to the grid, and reactive power is not shared by the PV array system [8-12]. If the load requires any reactive power, then the grid has to provide reactive load power. The reactive power compensation in the load side can be done by using a capacitor bank [13-17]. But reactive power compensation by fixed capacitor bank has some demerits such as reactive power supplied by the fixed capacitor is always constant based on installed capacitor bank in the load side, i.e., the requirement of the reactive power in the load side is less than the installed capacity of the capacitor bank then power factor in the load side goes to leading, and if reactive power requirement by the load is high than the installed capacity of the capacitor bank, then power factor in the load side is lagging. Due to this, the penalty will give to the consumer by the energy supplying company [17-22]. In this paper, STATCOM is presented for solar PV array integrated grid system to compensate the reactive power dynamically to overcome the problem in the fixed capacitor bank.

The paper is vv organized in the following way: Design of solar PV panel array, harmonic filter in the grid-tied inverter, DC-DC converter, the harmonic filter of the STATCOM side, and fixed capacitor bank is presented in section II. The simulation of solar PV array integrated grid system with different operating are explained in section III. In section IV provides the closing remarks of the proposed work.

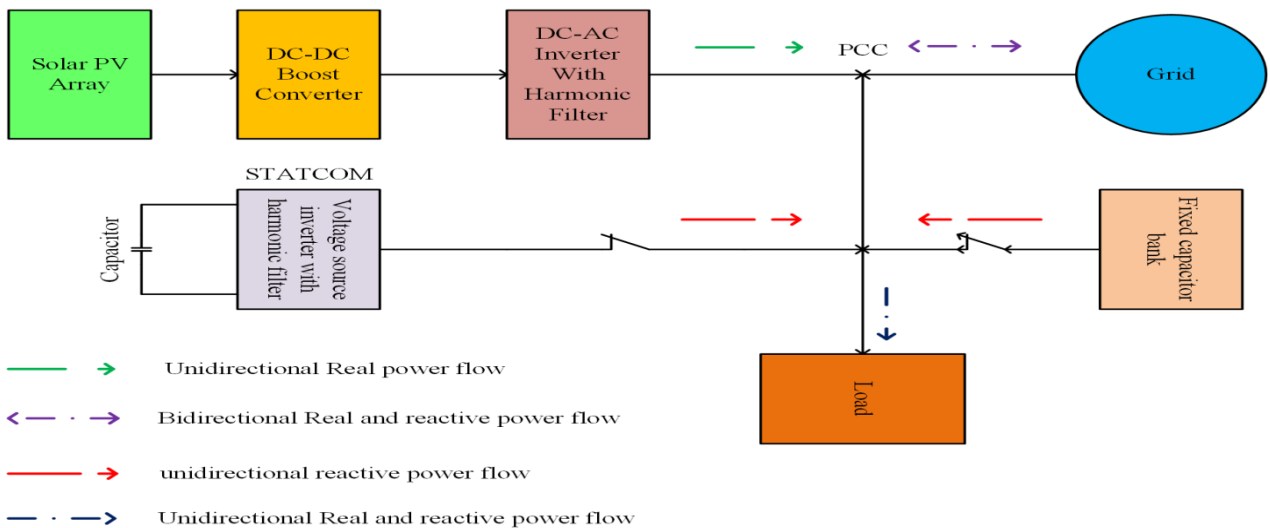

**Fig. 1Grid connected system with STATCOM and fixed capacitor.**

## II. DESIGN OF SOLAR PHOTOVOLTAIC GRID SYSTEM WITH STATCOM AND FIXED CAPACITOR

In this section, the design of the Photovoltaic grid-connected system, the design of STATCOM, and the fixed capacitor are explained. Fig.1 shown the functional diagram of the solar Photovoltaic grid system with STATCOM and fixed capacitor.

The system consists of a solar Photovoltaic array, grid-tied inverter for integration of solar Photovoltaic, DC-DC boost converter, LCL filter in the inverter of the solar PV side, STATCOM dc-link capacitor, voltage source inverter, LCL filter in the statcom side, and fixed capacitor. DC-DC converter used for integrating Solar PV array with the grid-tied inverter. A three-phase grid is integrated into a grid-tied inverter via LCL (Inductor – Capacitor - Inductor) filter. STATCOM, a three-phase Voltage source inverter, is integrated into the three-phase grid is connected via an LCL filter. The fixed capacitor is integrated directly into the three-phase grid via three circuit breakers.

### A. Design of Solar PV array system

The solar PV panel system can be formed through parallel and series integration of the different solar PV panels. The series connection of the solar PV panel will increase the solar PV panel array terminal voltage, and the parallel connection of the solar PV panel array will increase the solar PV panel array current rating. The rating of the single panel is 213.15 W, maximum power point PV panel voltage is 29 V, PV panel voltage at the open circuit is 36.3 V, PV current at the short circuit is 7.84 A, and maximum power point cutter is 7.35 A. series-connected PV panel is 10, and parallel-connected PV panel is 47. The overall rating of the solar PV array is 100.345 kW, PV panel voltage at the open circuit is 363 V, PV current at the short circuit is 368.48 A, maximum power point current is 345.45 A, and maximum power point voltage is 290 v. Fig.2 shows the power-voltage and current-voltage characteristics of the consider solar PV array for different irradiance conditions. The solar PV array has maximum power is 9.72 kW at 100 W/m² and 25°C, maximum power is 50.75 kW at 500 W/m² and 25°C and maximum power is 100.345 kW at 1000 W/m² and 25°C.

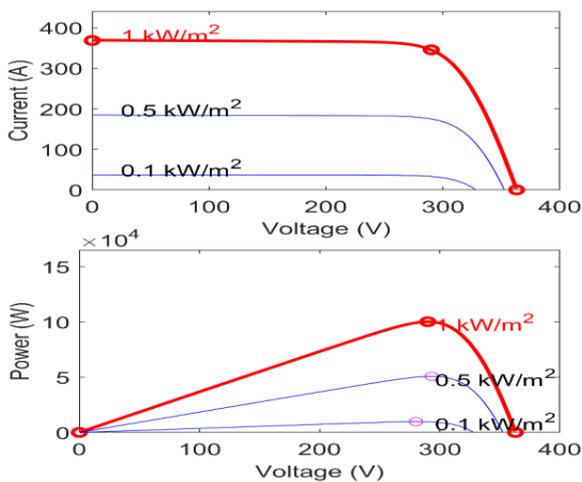

**Fig. 1 For varying irradiances, the P-V and I-V characteristics of the considered Solar PV array.**

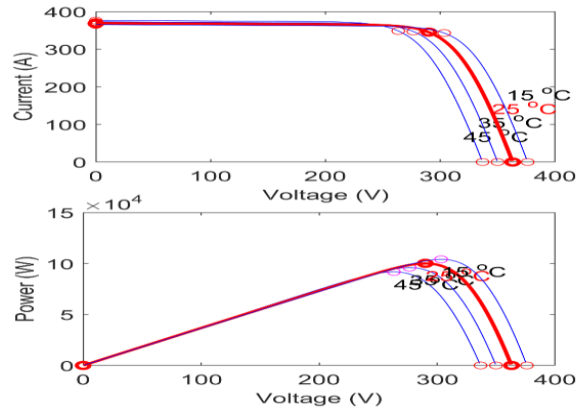

**Fig. 2 For varying temperatures, the P-V and I-V characteristics of the considered Solar PV array.**

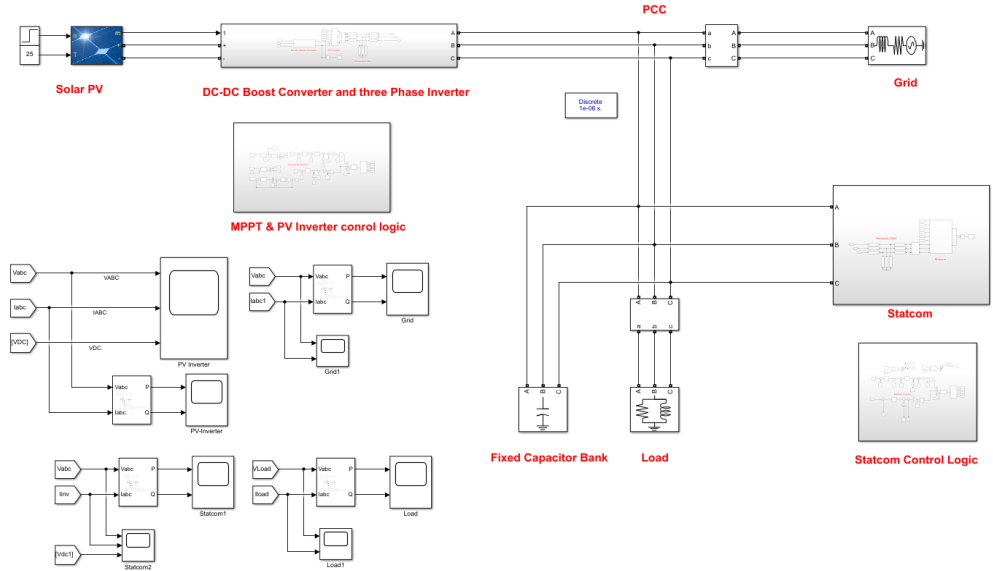

**Fig. 3 Simulink model of photovoltaic solar array integrated grid system with fixed capacitor bank and STATCOM.**

Figure 3 Depicts the P-V and I-V characteristics of the considered Solar PV under consideration for various temperatures. The solar PV has maximum power is 104.1 kW at 15°C and 1000 W/m², maximum power is 100.345 kW at 25°C and 1000 W/m², maximum power is 98.78 kW at 35°C and 1000 W/m², and maximum power is 95.03 kW at 45°C and 1000 W/m².

## B. Design of DC-DC Boost Converter

The direct current – direct current step-up converter is designed based on voltage and power ratings of the solar PV array and output voltage requirement of the grid-tied inverter. The standard testing condition ratings of the solar PV are 290 V and 100.345 kW, respectively. The output voltage requirement or DC link requirement for the grid-tied inverter is 700 V. the following equation used to design the inductor and capacitor of the direct current – direct current step-up converter in the solar PV system for grid integration [23],

$$I_{outmax} = \frac{P}{V_{out}} \qquad (1)$$

$$\Delta I_L = 0.07 \times I_{outmax} \times \frac{V_{out}}{V_{in}} \qquad (2)$$

$$\Delta V_{out} = 0.007 \times V_{out} \qquad (3)$$

$$L = \frac{V_{in} \times (V_{out} - V_{in})}{\Delta I_L \times F_S \times V_{out}} \qquad (4)$$

$$C = \frac{I_{outmax} \times (1 - \frac{V_{in}}{V_{out}})}{\Delta V_{out} \times F_S} \qquad (5)$$

Where L in the filter inductor of the direct current – direct current step-up converter, C is the filter capacitor of the direct current – direct current step-up converter, standard test condition power rating of solar PV is denoted as P and $V_{in}$ is PV voltage at standard test conditions, $I_{outmax}$ is the maximum current of the converter at the terminal of the direct current – direct current step-up converter, $V_{out}$ is the output voltage of the direct current – direct current step-up converter, $\Delta I_L$ is the ripple inductor current of the direct current – direct current step-up converter, $\Delta V_{out}$ is the ripple output voltage of the direct current – direct current step-up converter and $F_s$ is the switching frequency of the direct current – direct current step-up converter.

The maximum current of the direct current – direct current step-up converter at the output terminal is,

$$I_{outmax} = \frac{100.345 \times 1000}{700} = 143.35 \ A$$

The ripple inductor current of the direct current – direct current step-up converter is,

$$\Delta I_L = 0.07 \times 143.34 \times \frac{700}{290} = 24.21 \ A$$

The ripple inductor voltage of the direct current – direct current step-up converter is,

$$\Delta V_{out} = 0.007 \times 700 = 4.9 \ V$$

The designed value of the inductor of the direct current – direct current step-up converter is,

$$L = \frac{290 \times (700 - 290)}{24.21 \times 5000 \times 700} = 1.4 \ mH$$

The designed value of the capacitor of the direct current – direct current step-up converter is,

$$C = \frac{143.35 \times (1 - \frac{290}{700})}{4.9 \times 5000} = 3400 \ \mu F$$

## C. Grid-tied inverter harmonic Filter design

The harmonic filter for the grid-tied inverter is designed based on solar power ratings, switching frequency of the inverter, dc-link voltage, grid voltage, and grid frequency. The following equation used for the design of harmonic LCL filter for grid-tied inverter [24-25],

$$\omega_g = 2 \times \pi \times f_g \qquad (6)$$

$$Z_b = \frac{V_g^2}{\left(\frac{P}{3}\right)} \qquad (7)$$

$$C_b = \frac{1}{\omega_g \times Z_b} \qquad (8)$$

$$I_{max} = \frac{P \times \sqrt{2}}{3 \times V_g \times 0.9} \qquad (9)$$

$$\Delta I_{max} = \frac{10}{100} \times I_{max} \qquad (10)$$

$$L_1 = \frac{V_{out}}{6 \times F_{sw} \times \Delta I_{max}} \qquad (11)$$

$$C_g = 0.05 \times C_b \qquad (12)$$

$$L_2 = \frac{\sqrt{\frac{1}{0.2^2}} + 1}{C_g \times (2 \times \pi \times F_{sw})^2} \qquad (13)$$

$$\omega_{res} = \sqrt{\frac{L_1 + L_2}{L_1 \times L_2 \times C_g}} \qquad (14)$$

$$\omega_{min} = 10 \times f_g \qquad (15)$$

$$\omega_{max} = 0.5 \times F_{sw} \qquad (16)$$

Where, $f_g$ is the grid frequency, $V_g$ is the grid voltage per phase, $\omega_g$ angular frequency in radians of the grid, $Z_b$ is per phase impedance of the harmonic filter, $C_b$ is the harmonic filter base capacitance, $I_{max}$ is the harmonic filter maximum current rating, $\Delta I_{max}$ is the harmonic filter maximum current ripple of the filter inductor, $L_1$ is the input side filter inductor of the harmonic filter, $L_2$ is harmonic filter output side filter inductor, $C_g$ is harmonic filter capacitor of the, $\omega_{res}$ is the resonance frequency of harmonic filter in radians, $\omega_{min}$ is the minimum frequency limit of the harmonic filter and $\omega_{max}$ is the maximum frequency limit of the harmonic filter.

The angular frequency of the grid system is given by,
$$\omega_g = 2 \times \pi \times 50 = 314.1593 \; rad/s$$

The per impedance of the harmonic filter is given by,
$$Z_b = \frac{230^2}{\left(\frac{100.345 \times 1000}{3}\right)} = 1.5870$$

The base capacitance of the harmonic filter is given by,
$$C_b = \frac{1}{314.1593 \times 1.5870} = 0.002$$

The maximum current rating of the harmonic filter is given by,
$$I_{max} = \frac{100.345 \times 1000 \times \sqrt{2}}{3 \times 230 \times 0.9} = 227.7317 \; A$$

The maximum ripple inductor of the harmonic filter is given by,
$$\Delta I_{max} = \frac{10}{100} \times 227.7317 = 22.7732 \; A$$

The input side filter inductor of the harmonic filter is given by,
$$L_1 = \frac{700}{6 \times 10000 \times 22.7732} = 0.6 \; mH$$

The filter capacitor of the harmonic filter is given by,
$$C_g = 0.05 \times 0.002 = 100.29 \; \mu F$$

The outside filter inductor of the harmonic filter is given by,

$$L_2 = \frac{\sqrt{\frac{1}{0.2^2}} + 1}{100.29 \times 10^{-6} \times (2 \times \pi \times 10000)^2} = 15 \; \mu H$$

### D. Design of harmonic Filter for STATCOM

The harmonic filter for STATCOM is designed based on the reactive power requirement, dc-link voltage, grid voltage, the switching frequency of the inverter, and grid frequency. A similar equation from (6) to (16) is followed for the design of the harmonic filter of the STATCOM. The designed value of the input side inductor of the harmonic filter in the STATCOM is 0.6 mH, the designed value of the output side inductor of the harmonic filter in the STATCOM is 15.15 μH, and the designed value of the capacitor of the harmonic filter in the STATCOM is 100.29 μF.

### E. Design of Fixed capacitor bank ratings

The reactive power rating of the fixed capacitor bank is determined by the load side's reactive power requirement. A 100 kVAr reactive power capacitor bank is used in this work for reactive power compensation.

## III. SIMULATION RESULTS AND DISCUSSION

In this section, investigation of reactive power compensation for a PV grid system with STATCOM and a fixed capacitor bank under various operating conditions. Figure 4 depicts the overall Simulink model of the proposed PV grid system.

The system suggested was tested under the following operating conditions, grid-connected solar PV array system supplying power to the load locally, irradiance variation, and reactive power compensation in a PV grid system along with a fixed capacitor bank, and reactive power compensation in a PV array integrated grid system with a STATCOM.

### A. Simulation results of the PV integrated grid system with supplying only real power into the grid and supplying local load(Case 1)

In this condition, local load real power is set at 50 kW, and reactive power is set at 100 kvar. The solar PV array irradiance change from 1000 w/m² to 500 W/m² at 0.1 sec.

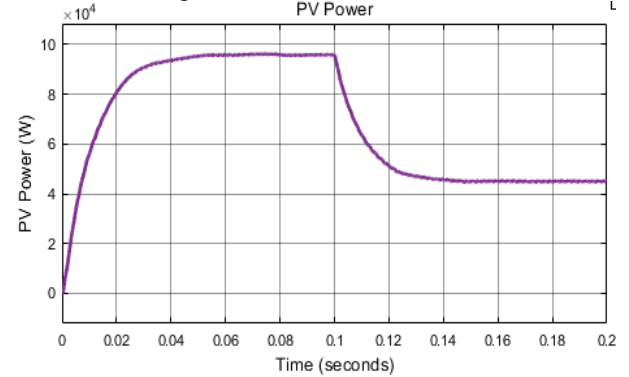

**Fig. 4 Solar PV array power for case 1.**

Fig.5 shows the solar PV array power variation of a solar PV array as the irradiance changes from 1000 W/m² to 500 W/m² over 0.1 seconds. The maximum power of solar PV panels at 1000 W/m² is 95.61 kW and at 500 W/m² is 44.86 kW. The solar PV inverter's reactive and real power is depicted in Fig.6. The PV inverter voltage, inverter current, and DC link voltage are shown in Fig.7.

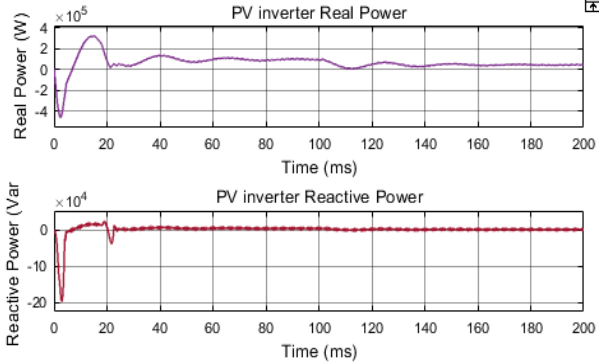

**Fig. 5 Solar PV inverter real reactive power for case 1.**

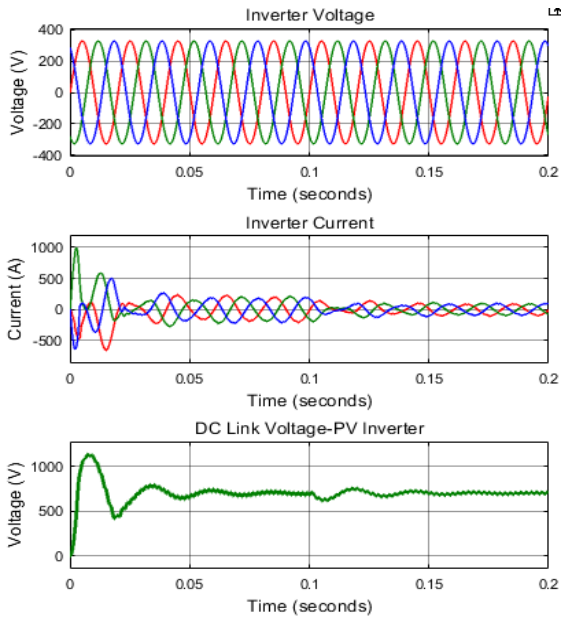

**Fig. 6 PV inverter dc-link voltage, voltage, and current for case 1.**

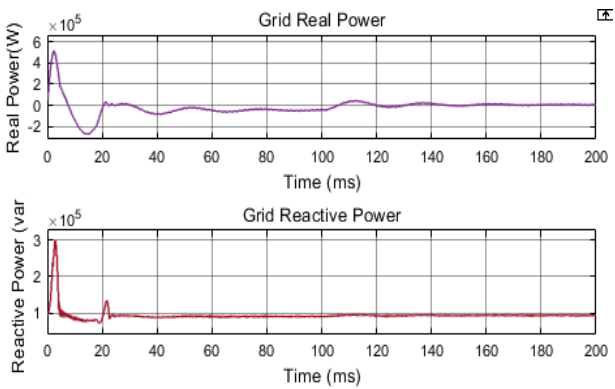

**Fig. 7 Grid reactive and real power for case 1.**

The grid's reactive and real power is depicted in Fig.8. the grid's real power is -44.8 kW during 1000 W/m² irradiance i.e., the grid receiving real power from the solar PV inverter and 892 W during 500 W/m² irradiances i.e., grid supplying real power to load. The reactive power of the grid is 100.1 kvar. The grid voltage and grid current is depicted in Fig.9.

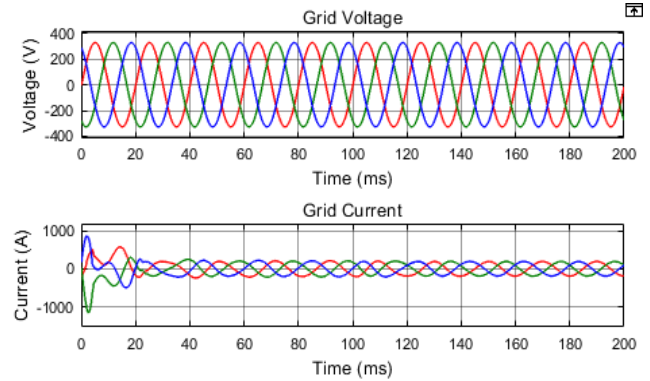

**Fig. 8 Grid current and grid voltage for case 1.**

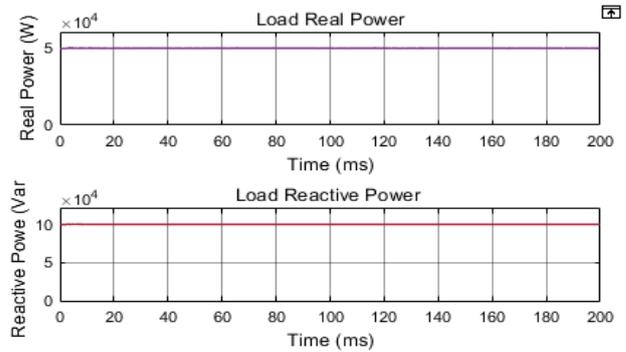

**Fig. 9 Load reactive and real power for case 1.**

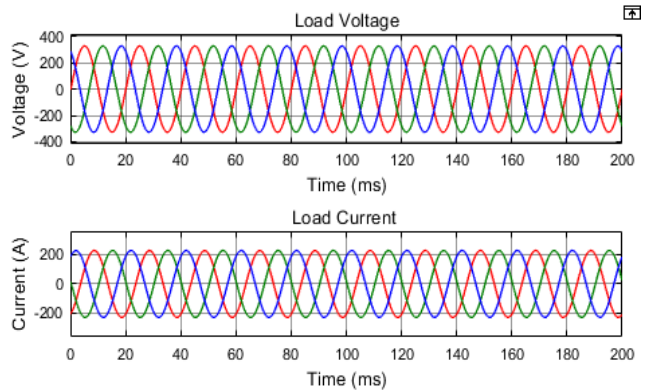

**Fig. 10  Load voltage and current for case 1.**

The load reactive power and real power are depicted in Fig.10. the load real power is 50 kW during 1000 W/m² irradiance and 50 kW during 500 W/m² irradiances. The load reactive power is around 100 kVAr. The load current and load voltage is depicted in Fig.11.

### B.  Simulation results of the solar PV array integrated grid system with Fixed capacitor bank (Case 2)

In this condition, local load real power is set at 100 kW, and reactive power is set at 150 kVAr. The PV array irradiance change from 500 w/m² to 1000 W/m² at 0.1 sec. Fig.12 shows the variation of solar PV array power for case 2. the solar PV panel's maximum power output at 500 W/m² is 44.97 kW and the solar PV panel's maximum power output is 96.26 kW at 1000 W/m². The solar PV inverter reactive power and real power are depicted in Fig.13. the solar PV inverter's real power is 44.12 kW during 500 W/m² irradiances and 95.96 kW during 1000 W/m² irradiance. The PV inverter dc-link voltage, voltage, and current is depicted in Fig.14.

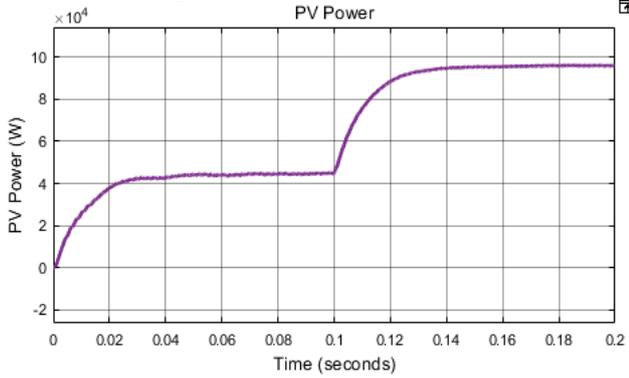

**Fig. 11 Solar PV array power for case 2.**

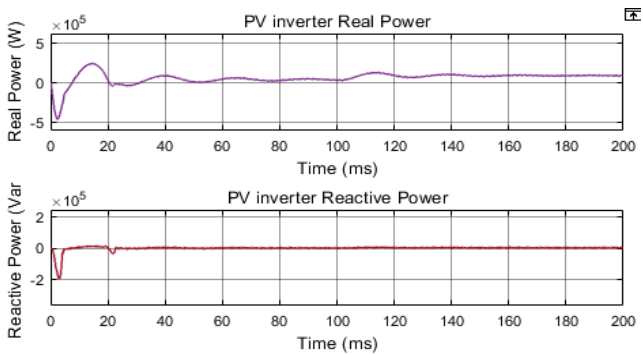

**Fig. 12 PV inverter reactive power and real for case 2.**

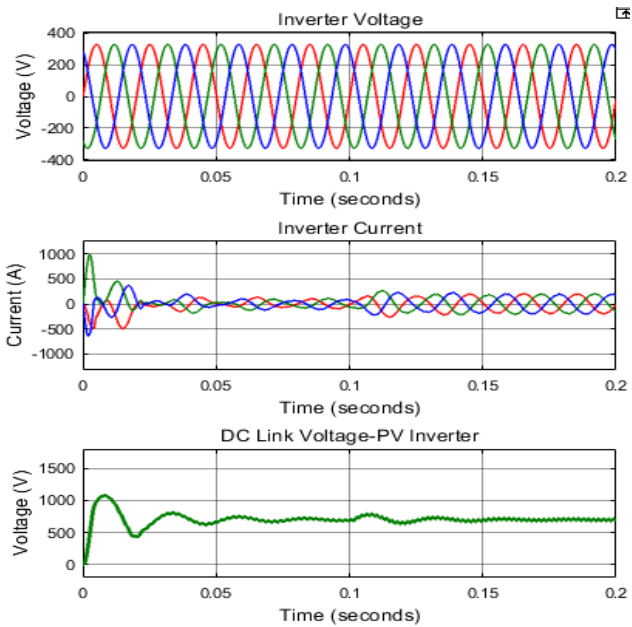

**Fig. 13 PV inverter voltage and current for case 2.**

The grid reactive power and real power are depicted in Fig.15. the grid real power is 44.8 kW during 500 W/m² irradiances i.e., grid supplying real power to the load, and 506W during 1000 W/m² irradiance i.e., load also receiving power from the grid. The reactive power sourced by the grid is 1.2kvar. The grid current and voltage are depicted in Fig.16.

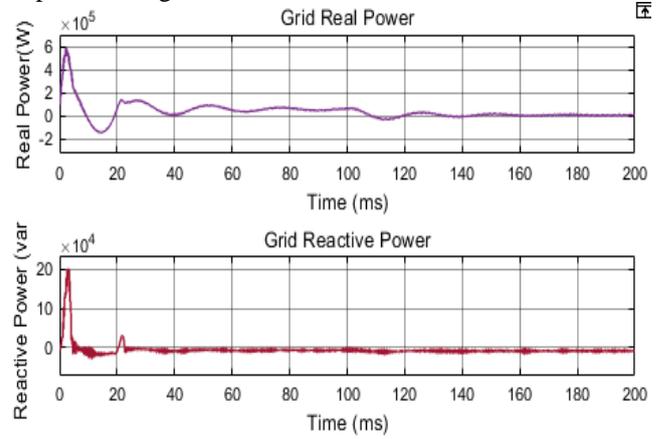

**Fig. 14 Grid reactive power and real power for case 2.**

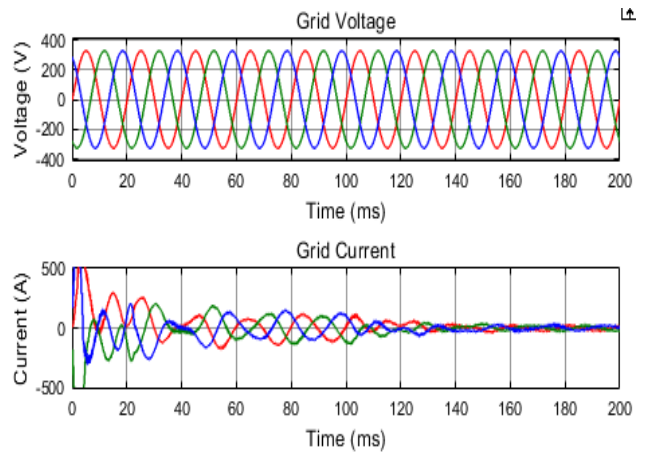

**Fig. 15 Grid current and grid voltage for case 2.**

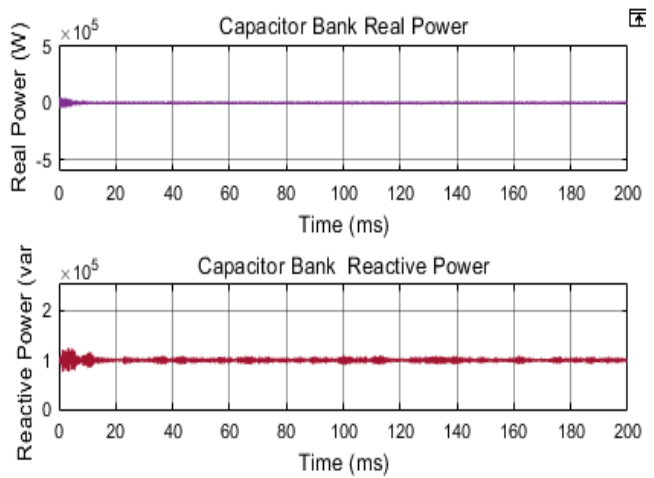

**Fig. 16 Capacitor back reactive power and real power for case 2.**

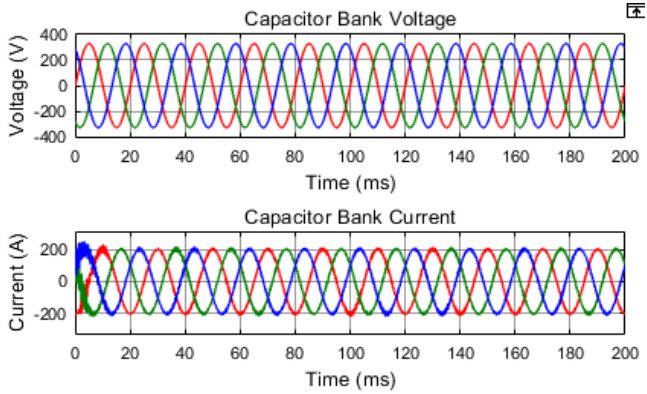

**Fig. 17 Capacitor bank voltage and current for case 2.**

The capacitor bank reactive power and real power and is depicted in Fig.17. the capacitor bank real power is 1.3 kW. The capacitor bank supplying reactive power to load is around 100kVAr. The capacitor bank voltage, the current is shown in Fig.18.

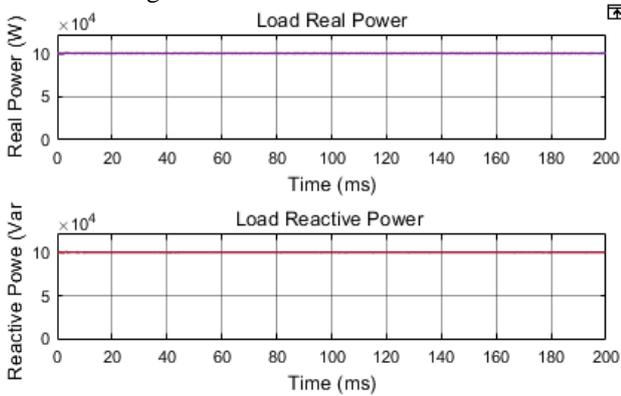

**Fig. 18 Load reactive power and real power for case 2.**

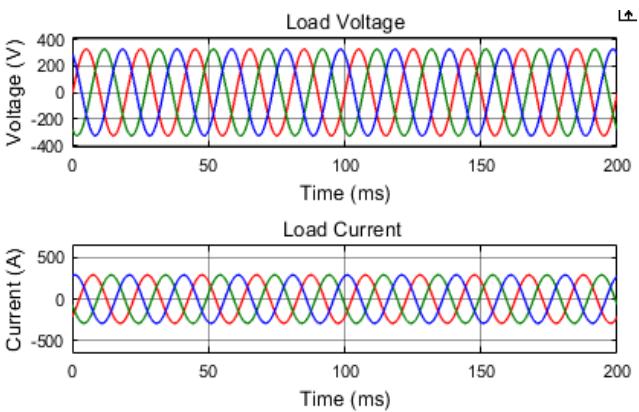

**Fig. 19  Load current and voltage for case 2**.

The load real power and reactive power are shown in Fig.19. the load real power is 100 kW during 1000 W/m$^2$ irradiance and 100 kW during 500 W/m$^2$ irradiances. The reactive power of the load is 100 kVAr. The load voltage, the current are displayed in Fig.20.

## C. Simulation results of the PV integrated grid system with STATCOM (Case 3)

In this condition, local load real power is set at 10 kW, and reactive power is set at 150 kVAr. The solar PV array irradiance change from 500 w/m$^2$ to 1000 W/m$^2$ at 0.1 sec.

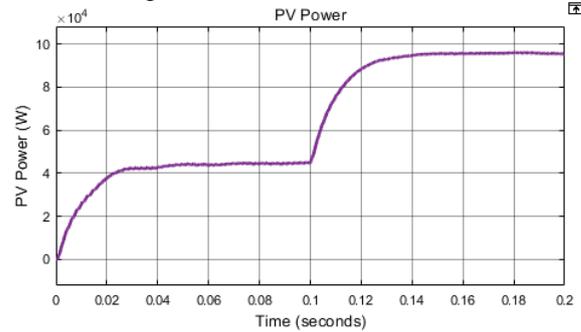

**Fig. 20 Solar PV array power for case 3.**

Fig.21 shows the solar PV array power variation for irradiance change from 500 W/m$^2$ to 1000 W/m$^2$ at 0.1 sec. the solar PV array maximum power at 500 W/m$^2$ is 44.99 kW and the solar PV array maximum power is 96.3 kW at 1000 W/m$^2$. The solar PV inverter reactive power and real power are depicted in Fig.22. the solar PV inverter's real power is 44.15 kW during 500 W/m$^2$ irradiances and 95.99 kW during 1000 W/m$^2$ irradiance. The PV inverter dc-link voltage, voltage, and current are seen in Fig.23.

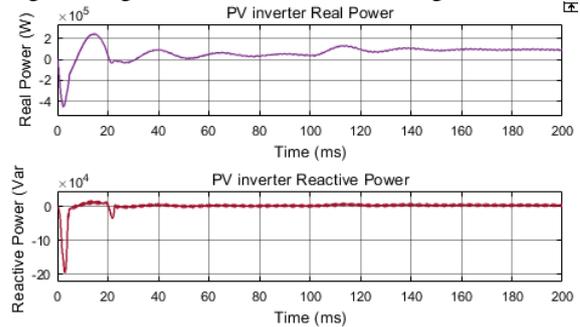

**Fig. 21 PV inverter reactive power and real power for case 3.**

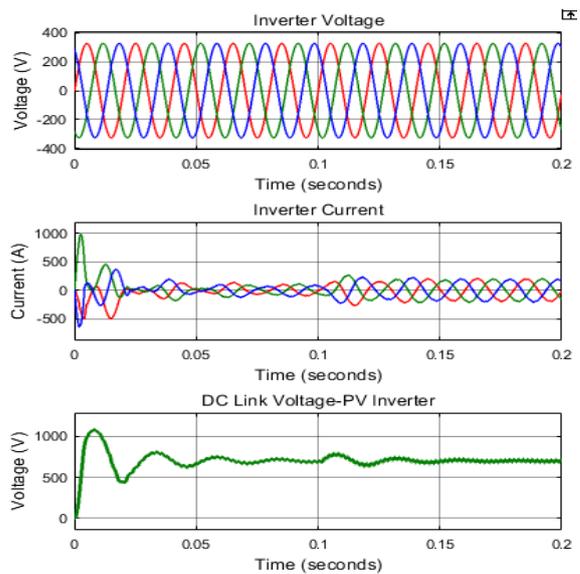

**Fig. 22 PV inverter voltage and current for case 3.**

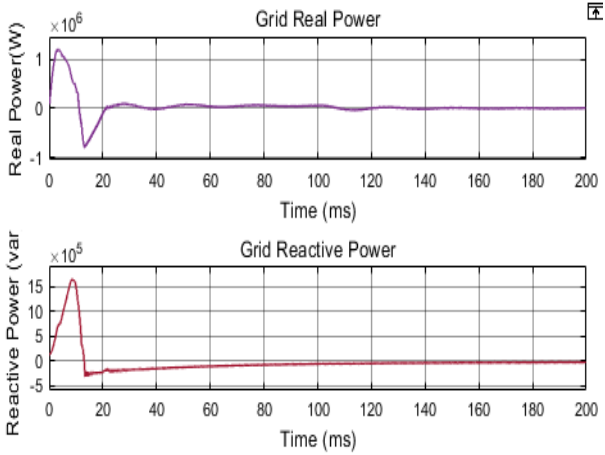

**Fig. 23 Grid reactive power and real power for case 3.**

The grid reactive power and real power are seen in Fig.24. The grid real power is 1.3 kW during 500 W/m² irradiances i.e., grid supplying real power to the load. The grid's reactive power is 1.2 kVAr. The grid current and voltage are depicted in Fig.25.

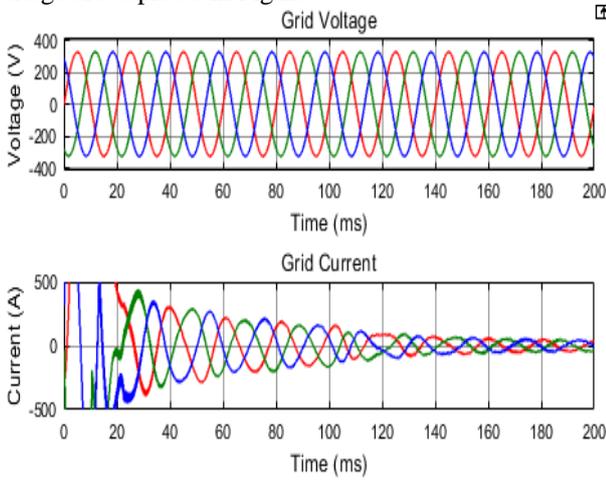

**Fig. 24 Grid current and voltage for case 3.**

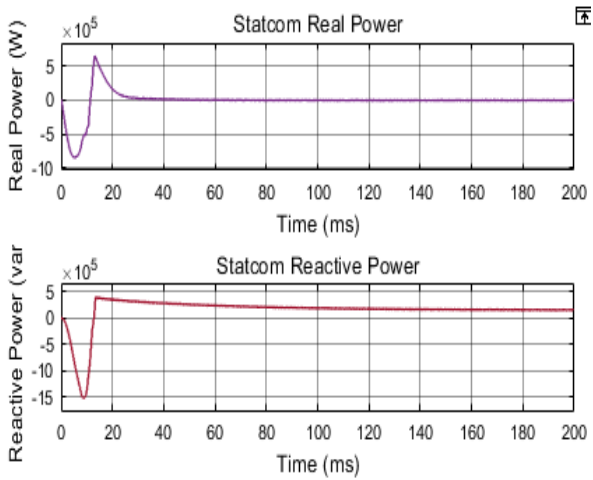

**Fig. 25 STATCOM reactive power and real power for case 3.**

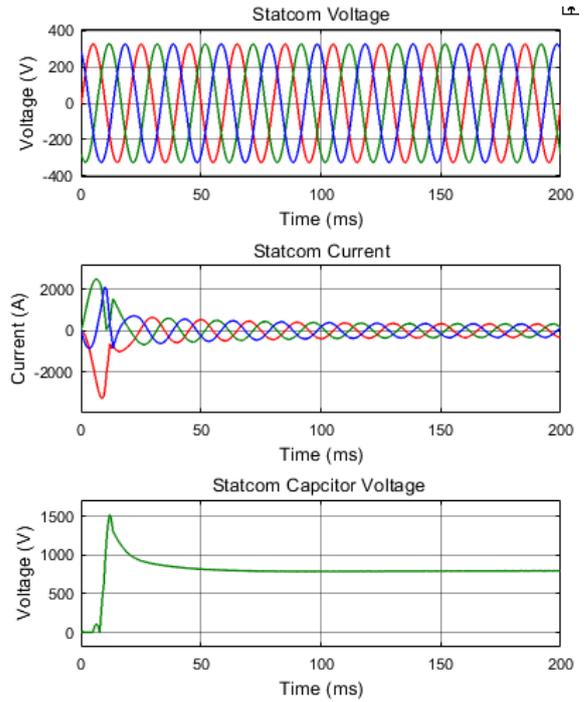

**Fig. 26 STATCOM dc-link voltage, voltage, and current for case 3.**

The STATCOM reactive power and real power are depicted in Fig.26. The STATCOM real power is 800 w. The reactive power of the STATCOM is 150.2kvar. The STATCOM voltage, current, and dc-link voltage are shown in Fig.27.

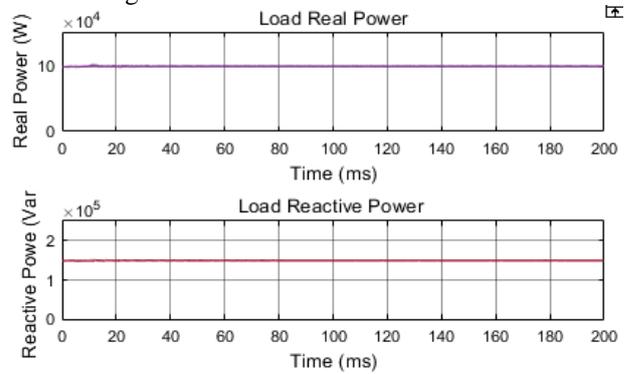

**Fig. 27 Load reactive power and real for case 3.**

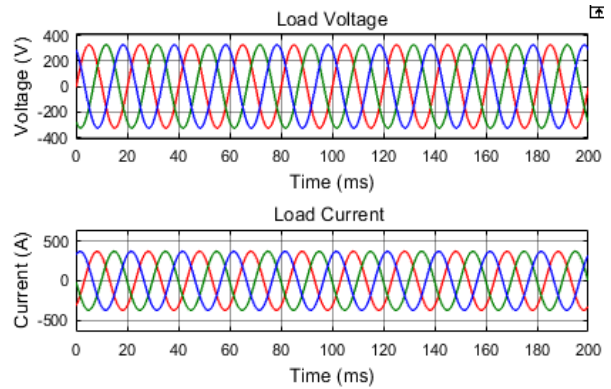

**Fig. 28 Load current and voltage for case 3.**

The load reactive power and real power are depicted in Fig.28. the load real power is 100 kW during 1000 W/m² irradiance and 100 kW during 500 W/m² irradiances. The load reactive power is around 150 kVAr. The load current and voltage are depicted in Fig.29.

## IV. CONCLUSIONS

This study shows how to use STATCOM and a fixed capacitor bank to compensate for reactive power in a grid-connected solar PV array. This article explains how to design a solar PV system, a DC-DC boost converter, a grid-tied inverter, and a STATCOM voltage source inverter. The suggested system's whole model is simulated using the MATLAB Simulink software suite. The grid-connected photovoltaic solar array system has been tested under different operating conditions, without fixed capacitor and STATCOM and fixed capacitor and STATCOM for reactive power compensation. From the simulation results, reactive power of the grid-connected solar PV array system is compensated effectively and dynamically by STATCOM based on reactive power requirement in the load side. In the grid-connected solar PV array system, STATCOM's performance in reactive power compensation is better than that of fixed capacitor banks.